\documentclass{article}
\usepackage{longtable}
\usepackage{float}
\usepackage{subfloat}
\usepackage{lscape,epsfig}
\usepackage{graphicx}
\usepackage{amssymb}
\usepackage{amsmath}
\usepackage{natbib}
\textheight=22cm \DeclareSymbolFont{ppa}{OT1}{ppl}{m}{it}
\DeclareMathSymbol{\vv}{\mathalpha}{ppa}{'166}

minus 2.3mu \thickmuskip = 2.6mu plus 2mu minus 2.6mu

\let\svthefootnote\thefootnote

\begin{document}

\newcommand{\dd}{\,{\rm d}}
\newcommand{\ie}{{\it i.e.},\,}
\newcommand{\etal}{{\it $et$ $al$.\ }}
\newcommand{\eg}{{\it e.g.},\,}
\newcommand{\cf}{{\it cf.\ }}
\newcommand{\vs}{{\it vs.\ }}
\newcommand{\zdot}{\makebox[0pt][l]{.}}
\newcommand{\up}[1]{\ifmmode^{\rm #1}\else$^{\rm #1}$\fi}
\newcommand{\dn}[1]{\ifmmode_{\rm #1}\else$_{\rm #1}$\fi}
\newcommand{\upd}{\up{d}}
\newcommand{\uph}{\up{h}}
\newcommand{\upm}{\up{m}}
\newcommand{\ups}{\up{s}}
\newcommand{\arcd}{\ifmmode^{\circ}\else$^{\circ}$\fi}
\newcommand{\arcm}{\ifmmode{'}\else$'$\fi}
\newcommand{\arcs}{\ifmmode{''}\else$''$\fi}
\newcommand{\MS}{{\rm M}\ifmmode_{\odot}\else$_{\odot}$\fi}
\newcommand{\RS}{{\rm R}\ifmmode_{\odot}\else$_{\odot}$\fi}
\newcommand{\LS}{{\rm L}\ifmmode_{\odot}\else$_{\odot}$\fi}

\newcommand{\Abstract}[2]{{\footnotesize\begin{center}ABSTRACT\end{center}
\vspace{1mm}\par#1\par \noindent {~}{\it #2}}}

\newcommand{\TabCap}[2]{\begin{center}\parbox[t]{#1}{\begin{center}
  \small {\spaceskip 2pt plus 1pt minus 1pt T a b l e}
  \refstepcounter{table}\thetable \\[2mm]
  \footnotesize #2 \end{center}}\end{center}}

\newcommand{\TableSep}[2]{\begin{table}[p]\vspace{#1}
\TabCap{#2}\end{table}}

\newcommand{\FigCap}[1]{\footnotesize\par\noindent Fig.\  %
  \refstepcounter{figure}\thefigure. #1\par}

\newcommand{\TableFont}{\footnotesize}
\newcommand{\TableFontIt}{\ttit}
\newcommand{\SetTableFont}[1]{\renewcommand{\TableFont}{#1}}
\newcommand{\MakeTable}[4]{\begin{table}[htb]\TabCap{#2}{#3}
  \begin{center} \TableFont \begin{tabular}{#1} #4
  \end{tabular}\end{center}\end{table}}

\newcommand{\MakeTableSep}[4]{\begin{table}[p]\TabCap{#2}{#3}
  \begin{center} \TableFont \begin{tabular}{#1} #4
  \end{tabular}\end{center}\end{table}}

\newenvironment{references}%
{ \footnotesize \frenchspacing
\renewcommand{\thesection}{}
\renewcommand{\in}{{\rm in }}
\renewcommand{\AA}{Astron.\ Astrophys.}
\newcommand{\AAS}{Astron.~Astrophys.~Suppl.~Ser.}
\newcommand{\ApJ}{Astrophys.\ J.}
\newcommand{\ApJS}{Astrophys.\ J.~Suppl.~Ser.}
\newcommand{\ApJL}{Astrophys.\ J.~Letters}
\newcommand{\AJ}{Astron.\ J.}
\newcommand{\IBVS}{IBVS}
\newcommand{\PASP}{P.A.S.P.}
\newcommand{\Acta}{Acta Astron.}
\newcommand{\MNRAS}{MNRAS}
\renewcommand{\and}{{\rm and }}
\section{{\rm REFERENCES}}
\sloppy \hyphenpenalty10000
\begin{list}{}{\leftmargin1cm\listparindent-1cm
\itemindent\listparindent\parsep0pt\itemsep0pt}}%
{\end{list}\vspace{2mm}}

\def\TYLDA{~}
\newlength{\DW}
\settowidth{\DW}{0}
\newcommand{\dw}{\hspace{\DW}}

\newcommand{\refitem}[5]{\item[]{#1} #2%
\def\REFARG{#3}\ifx\REFARG\TYLDA\else, {\it#3}\fi
\def\REFARG{#4}\ifx\REFARG\TYLDA\else, {\bf#4}\fi
\def\REFARG{#5}\ifx\REFARG\TYLDA\else, {#5}\fi.}

\newcommand{\Section}[1]{\section{#1}}
\newcommand{\Subsection}[1]{\subsection{#1}}
\newcommand{\Acknow}[1]{\par\vspace{5mm}{\bf Acknowledgements.} #1}
\pagestyle{myheadings}

\newfont{\bb}{ptmbi8t at 12pt}
\newcommand{\xrule}{\rule{0pt}{2.5ex}}
\newcommand{\xxrule}{\rule[-1.8ex]{0pt}{4.5ex}}

\begin{center}
{\Large\bf
 The Clusters AgeS Experiment (CASE).\dag  \\
 Variable stars in the field of  
 the globular cluster NGC~362}{\LARGE$^\ast$}
 \vskip1cm
  {\large
      ~~M.~~R~o~z~y~c~z~k~a$^1$,
      ~~I.~B.~~T~h~o~m~p~s~o~n$^2$,
      ~~W.~~N~a~r~l~o~c~h$^1$
      ~~W.~~P~y~c~h$^1$
      ~~and~~A.~~S~c~h~w~a~r~z~e~n~b~e~r~g~~-~~C~z~e~r~n~y$^1$
   }
  \vskip3mm
{ $^1$Nicolaus Copernicus Astronomical Center, ul. Bartycka 18, 00-716 Warsaw, Poland\\
     e-mail: (mnr, wnarloch, pych, alex)@camk.edu.pl\\
  $^2$The Observatories of the Carnegie Institution for Science, 813 Santa Barbara
      Street, Pasadena, CA 91101, USA\\
     e-mail: ian@obs.carnegiescience.edu}
\end{center}

\vspace*{7pt}
\Abstract 
{The field of the globular cluster NGC~362 was monitored between 1997 and 2015 in a search 
for variable stars. $BV$ light curves were obtained for 151 periodic or likely periodic 
variables, over a hundred of which are new detections. Twelve newly detected variables are 
proper-motion members of the cluster: two SX~Phe and two RR Lyr pulsators, one contact binary, 
three detached or semi-detached eclipsing binaries, and four spotted variables. The most 
interesting objects among these are the binary blue straggler V20 with an asymmetric
light curve, and the 8.1 d semidetached binary V24 located on the red giant branch of NGC~362, 
which is a Chandra X-ray source. We also provide substantial new data for 24 previously known 
variables. 
}
{globular clusters: individual (NGC~362) -- stars: variables -- 
stars: SX Phe -- blue stragglers -- binaries: eclipsing
}

\let\thefootnote\relax\footnotetext{\dag The project CASE was initiated and for long time led
by our friend and tutor Janusz Kaluzny, who prematurely passed away in March 2015.}
\let\thefootnote\relax\footnotetext
{$^{\mathrm{\ast}}$Based on data obtained with Swope and du Pont telescopes at the Las Campanas 
Observatory.}
\let\thefootnote\svthefootnote

\Section{Introduction} 
\label{sec:intro}
The core-collapse globular cluster NGC~362 is projected against the outskirts of the Small 
Magellanic Cloud (SMC) at $l=301^\circ.5$, $b=-46^\circ.3$, in a field with a low reddening 
of $E(B-V)=0.05$ mag. Its core radius $r_c$, half-light radius $r_h$, tidal radius $r_t$, [Fe/H] 
index, radial velocity, heliocentric distance $d_\odot$ and galactocentric distance $d_G$ are 
equal to 0$'$.18, 0$'$.82, 10$'$.4, 
-1.26, +223.5$\pm$0.5 km/s, 8.6 kpc and 9.4 kpc, respectively (Harris 1996, 2010 edition). 
The age of the cluster 
is estimated at $12.5\pm0.5$ Gyr, based 
on color-magnitude diagram (CMD) fitting with theoretical isochrones 
(Dotter et al. 2010), and its kinematics suggest an extragalactic origin (Forbes \& Bridges 2010). 
With an exceptionally red horizontal branch and large concentration parameter ($c=\log r_t/
r_c = 1.76$) NGC~362 may be regarded as a prime example of the correlation between horizontal 
branch morphology and central luminosity density observed among globular clusters (Dotter et al. 2010).

The cluster is a challenging target 
for studies with ground-based telescopes because of its distance from the Sun and high concentration.  The few pre-CCD searches for variables in the field of 
NGC~362, summarized by Clement et al. (2001; 2012 edition\footnote{
http://www.astro.utoronto.ca/~cclement/cat/C0100m711}) (hereafter C1-12), 
resulted in the detection of 16 objects. The two CCD surveys that have been 
performed so far (Sz\'ekely et al. 2007; hereafter 
Sz07, and Lebzelter \& Wood 2011; hereafter LW11) resulted in an 
additional 94 discoveries. In the whole
sample of 110 variables, 56 cluster members were found by C1-12, 
including 35 RR Lyr and four SX Phe 
pulsators, three eclipsing binaries, and 14 long-period/irregular variables. 

The 16 new variables belonging or probably belonging to NGC 362, and 88 new field variables presented 
in this contribution, are a result of the long-term photometric 
survey conducted within the CASE project (Kaluzny et al. 2005) using telescopes of the 
Las Campanas Observatory. Section~2 contains a brief report on the observations and explains the 
methods used to calibrate the photometry. Newly discovered variables are presented 
and discussed in Section~3. Section~4 contains new data on previously known variables which we 
consider worthy of publishing, and the paper is summarized in Section~5.
\section{Observations}
\label{sec:obs}
Our paper is based on images acquired mainly with the 1.0-m Swope telescope and the 
$2048\times 3150$ SITe3 camera. The field of view was $14.8\times 22.8$ arcmin$^2$ 
at a scale of 0.435 arcsec/pixel. Observations were obtained on 205 nights from 
July 7, 1997 to October 23, 2009. The same set of filters was used for all observations. 
A total of 3785 $V$-band images and 1123 $B$-band images were selected for  
analysis. The seeing ranged from 
1$''$.14 to 4$''$.3 and 1$''$.28 to 4$''$.07 for $V$ and $B$, respectively, with median 
values of 1$''$.67 and 1$''$.8. We also used 73 $V$-frames acquired in 2015 on Swope 
at the same resolution as before and with the same set of filters, but with the new E2V camera, 
and 270 frames acquired between 2001 and 2007 on the 2.5-m du Pont telescope with a field of 
view $8.84\times8.84$ arcmin$^2$ at a resolution of 0.259 arcsec/pixel.

The photometry was performed using an image subtraction technique implemented in the 
DIAPL package.\footnote{Available from http://users.camk.edu.pl/pych/DIAPL/index.html} 
To reduce the effects of PSF variability, each frame was divided into 3$\times$2  
overlapping subframes. The reference frames were constructed by combining 7 images 
in $V$ and 6 in $B$ with an average seeing of 1.$''$26 and 1.$''$33, respectively.
The light curves derived with DIAPL were converted from differential counts to magnitudes 
based on profile photometry and aperture corrections determined separately for each 
subframe of the reference frames. To extract the 
profile photometry from reference images and to derive aperture corrections, the 
standard Daophot, Allstar and Daogrow (Stetson 1987, 1990) packages were used. 
Profile photometry was also extracted for each individual image, enabling useful 
photometric measurements of stars which were overexposed on the reference frames. 
\subsection{Calibration}

The Swope/E2V and du Pont frames were reduced and measured separately from the Swope/SITe3 frames, 
and the results were transformed to the SITe3 instrumental system. The photometric calibration 
was based on SITe3 observations of 17 standards from three Landolt fields 
(Landolt 1992). During the night of August 6/7, 2000 each field was observed several times, 
yielding a total of 39 $V$ and $B$ measurements at air massess $1.06 < X < 2.01$. Based 
on those measurements, the following transformations of the SITe3 photometry to the standard 
system were derived:
\begin{align}
  v - V &= -2.500(1) + 0.018(1)\times(B-V) \nonumber\\
  b - B &= -2.581(2) + 0.032(2)\times(B-V) \nonumber\\
  b - v &= -0.078(1) + 0.951(1)\times(B-V) \nonumber,
\end{align}
where lower case and capital letters denote instrumental and standard magnitudes, 
respectively. 

%\subsection{Crowding}

\begin{figure}[H]
   \centerline{\includegraphics[width=0.95\textwidth,
               bb = 24 272 564 698, clip]{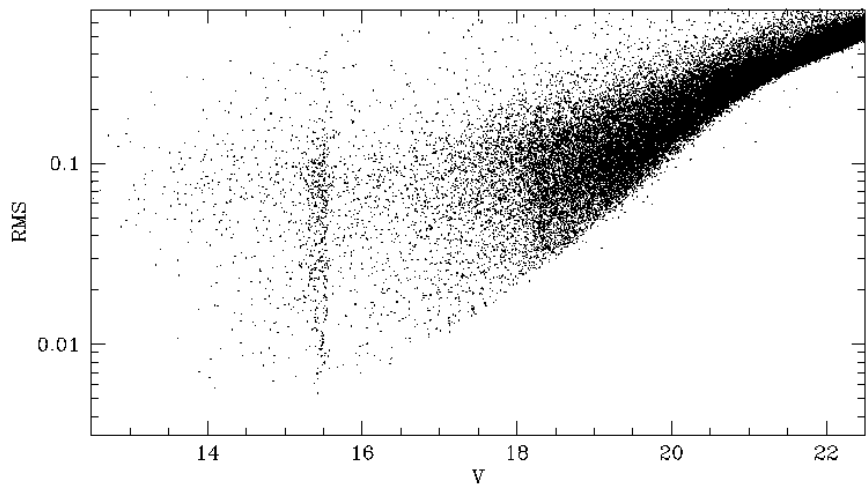}}
   \caption{ Standard deviation vs. average $V$ magnitude for
    light curves of stars from the NGC~362 field. 
    \label{fig:rms}}
\end{figure}

\begin{figure}[H]
   \centerline{\includegraphics[width=0.95\textwidth,
               bb = 55 200 562 687, clip]{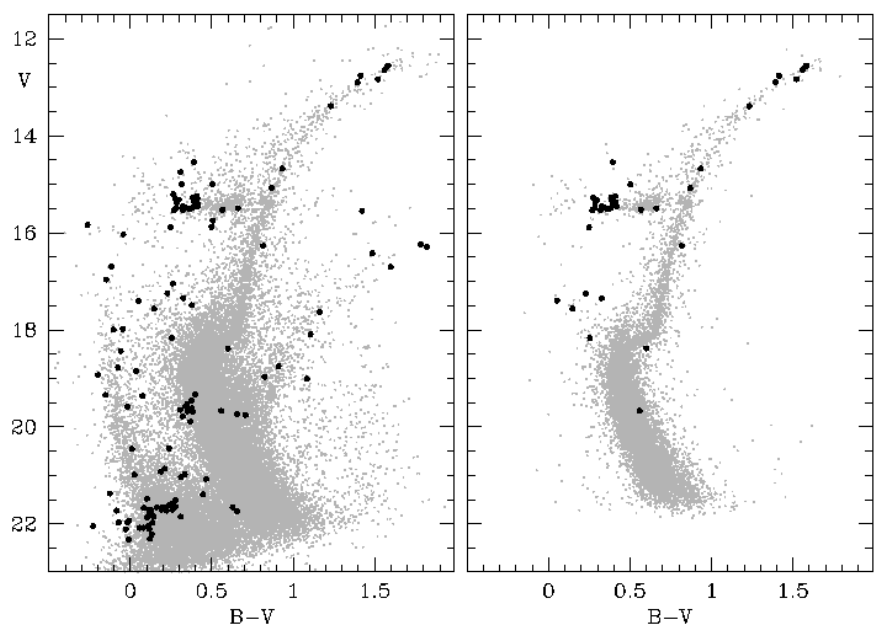}}
   \caption{CMD for the observed field. Left: all stars for which proper motions were measured. 
    Black points mark all the 138 variables detected within 
    the present survey for which  $B$-band magnitudes were available. Right: same as in the left 
    frame, but for the PM-members of the cluster only. 
    \label{fig:cmds}}
\end{figure}

Crowding in the field of view resulted in enhanced blending, which in turn significantly increased 
the scatter of the photometric measurements in the observed magnitude range. While for the much 
less concentrated NGC 3201 the smallest scatter at $V$ = 21 mag was 0.1 mag (Kaluzny et al. 2016), 
for NGC 362 it rose  to 0.25 mag (Fig. \ref{fig:rms}).
Fig.~\ref{fig:cmds}, based on the reference images, shows the CMD of the observed field. 
To make the figure readable, only stars with measured proper 
motions (Narloch et al., in preparation) are  selected to serve as a background for the 
variables. Stars identified  as proper-motion (PM) members of the cluster are shown in the right frame.
 
\subsection{Search for variables}

The search for variable stars was conducted using the AOV and AOVTRANS algorithms implemented in the 
TATRY code (Schwarzenberg-Czerny 1996 and 2012;  Schwar\-zenberg-Czerny \& Beaulieu 2006). We examined 
time-series photometric data of 51296 stars with $V<22$ mag. 
The scatter of photometric mesurements was significantly offset by the large number of available frames, 
and as a result we were able to detect small-amplitude pulsations of SX Phe/$\delta$ Sct stars from 
the SMC down to about $V=22.5$ mag. Light curves from both  telescopes were searched, but due 
to a much smaller number of frames  the data from duPont provide more information 
than those from Swope in only four cases. The du Pont data mainly served to make the finding charts. 

We obtained light curves for most of the previously  detected variables located  within 
our field of view\footnote{Available from the CASE archive at http://case.camk.edu.pl},
We discovered 102 variable or likely variable stars, 16 of which are PM-members or likely 
PM-members of NGC~362. Several RR Lyr stars belonging to the SMC were independently discovered by 
Soszy\'nski et al. (2016). 

\section{The new variables}

Basic data for the variables discovered within this survey are given in Tables \ref{tab:newvar}
and \ref{tab:fieldvar}. The equatorial coordinates conform to the UCAC4 system 
(Zacharias et al. 2013) and are accurate to 0$''$.2 - 0$''$.3. The $V$-magnitudes correspond to the 
maximum light in the case of eclipsing binaries; in the remaining cases average magnitudes are 
presented. 
Columns 5 --7 give $B-V$ color, amplitude in the $V$-band, and period of variability 
(an empty color entry means that no $B$-band photometry could be extracted). 
A CMD of the cluster with the locations of the variables identified is shown in Fig.~\ref{fig:cmd_var}. The
gray background stars are the PM-members of NGC~362 from the right frame of 
Fig.~\ref{fig:cmds}. Field objects are marked in black, those for which the PM data are missing or 
ambiguous in blue, and PM-members of the cluster in red.

\subsection {Members and possible members of NGC 362}
\label{sec:varmem}

\begin{figure}
    \centerline{\includegraphics[width=0.95\textwidth,
               bb = 146 312 870 731, clip]{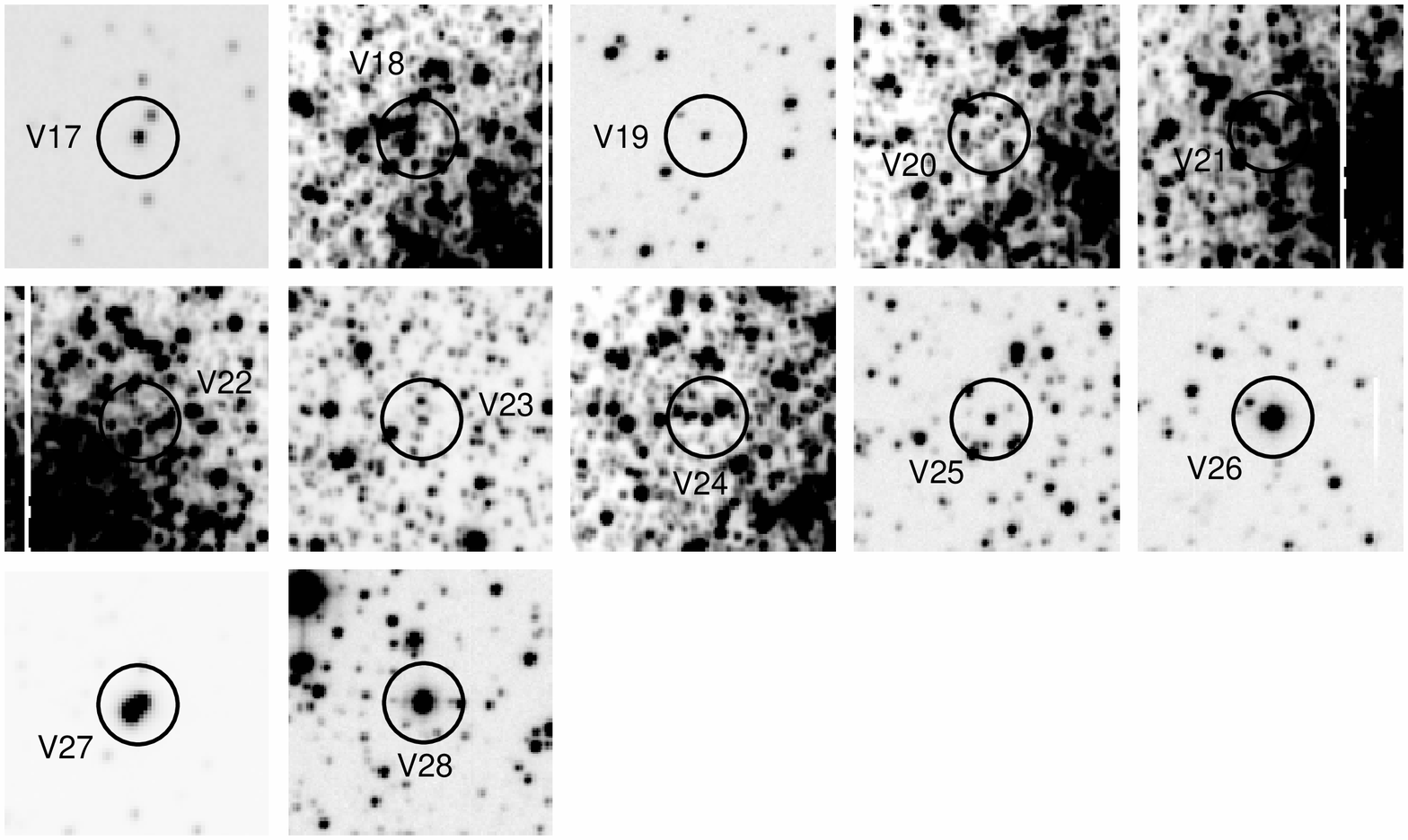}}
    \caption{Finding charts for the new variable members of NGC 362. Each 
     chart is 30$''$ on a side. North is up and East to the left.
     \label{fig:maps}}
\end{figure}

Fourteen of the newly detected periodic variables and two stars suspected of 
variability belong or likely belong to the cluster. The basic data for this 
sample are given in Table \ref{tab:newvar}. Here and in subsequent Tables,
magnitudes maximum light and average magnitudes are given for eclipsing binaries
and remaining variables, respectively, and amplitudes $\Delta V$ refer to directly 
observed curves (i.e. not to the phase-binned ones). To 
follow the naming convention of C1-12, we begin the numbering with V17. 
Entries in the last column 
of Table \ref{tab:newvar} indicate that stars V17 -- V28 are PM-members 
of NGC~362, whereas the 
membership status of VN01 -- VN04 is unclear. Figs.~\ref{fig:newvara} 
and \ref{fig:newvarb} show 
phased light curves of all the variables from Table~\ref{tab:newvar}; 
finding charts for the stars 
V17 -- V28 are presented in Fig.~\ref{fig:maps}.

V17 and V18 are SX Phe pulsators; V17 is also a blue straggler. Although we do not 
have  $B$-band data for V18, we are rather sure that it also must be located 
among the blue stragglers on the CMD. The large 
amplitude of this star, and the fact that our data seem to indicate 
multimode pulsations, make it a 
potential target for further studies. Well observed multimode pulsations 
would allow a determination of its 
mass, which in turn might provide valuable information about its origin. 

V19 is a typical W UMa-type binary with a total secondary eclipse, 
located about 1 mag below the turnoff. 
Although we were able to detect low amplitude pulsations of 
SX Phe/$\delta$ Sct from the SMC down to 
$V\approx22.5$ mag (see Table \ref{tab:SXSMC}), we did not 
find any W UMa stars fainter than V19. Thus, NGC 
362 provides another example of the paucity of contact binaries on 
the unevolved main sequences of 
globular clusters (Yan \& Mateo 1994; Kaluzny et al. 2014). 
At least in cluster environments, the 
principal factor enabling the formation of such systems from detached 
binaries seems to be
nuclear evolution: a contact configuration is achieved once the 
more massive component
approaches the turnoff and starts to expand. The frequently invoked 
alternative, magnetic braking
(e.g. St\c{e}pie\'n \& Gazeas 2012, and references therein), 
seems to play a minor role in the creation of W UMa stars.

V20 is an eclipsing blue straggler with an asymmetric light curve and 
broad minima differing by 0.65 mag in depth. From a detailed analysis 
of this system interesting conclusions concerning the evolution 
of blue stragglers might be obtained. Spectroscopic observations would 
be quite challenging given the 
orbital period of only 9.6 h, nevertheless it is worthwhile to take a 
spectrum just to check if lines 
of both components can be seen.

V21 and V22 are RRab pulsators residing in the core of NGC 362 at a projected distance $r_p\approx 
0.2r_c$ from its center. The light curve of V21 and its location about 1 mag above the horizontal 
branch are indicative of blending effects. In fact, in the archival HST frame 
10615\_01\_acs\_wfc\_f435w\_drz.fits there are two almost equally bright objects at the position 
of this star, separated by $\sim0''$.5. Another example of a tight pair of RR Lyrs is Sz31 described
in Sect. \ref{sec:oldvar} and in the Appendix. In that case the data were sufficient to disentangle 
the light curves; for V21 they are unfortunately too poor. The photometry of V22 is also severely 
affected by blending, but most probably there is  just one pulsating star in the composite
image.

Broad secondary minima identify V23 as a semidetached or nearly semidetached system. Such binaries
in globular clusters  are potentially interesting, however low brightness and blending make this one 
unsuitable 
for a detailed analysis (in the HST frame it is located $\sim0''$.5 from an equally bright object).

By far the most interesting object in Table \ref{tab:newvar} is the eclipsing binary V24, for which we 
have obtained a preliminary radial velocity curve. The nearly equal velocity amplitudes of 61.2 and 62.7 km/s 
and a period of 8.1 d identify this system as a pair of $\sim$0.8 M$_\odot$ stars. The shape of the 
light curve suggests a giant primary, most likely filling its Roche lobe, and a slightly hotter subgiant 
secondary. The light curve itself is extremely unstable -- its only stationary parts are ingress and 
egress of the primary minimum. 
The width of the secondary minimum varies by a factor of three on a timescale of a few years, suggesting
a variable flow of material between the components (perhaps an accretion disk is occasionally formed). 
The system coincides with the Chandra source X010322.09-705044.7, lending support to the mass 
transfer hypothesis. Unfortunately, the same properties that make V24 so interesting cause it to be 
a hard observational challenge, as several orbital periods of this system would have to be covered 
with high-quality photometry and spectroscopy in order to acquire a consistent picture of the mass flow 
and to obtain meaningful quantitative information about general physical parameters of the system.

Stars V26 and V28 belong to the red horizontal branch of NGC 362; most likely so do V25 and V27
for which we do not have $B$-band data. The low amplitude, slightly irregular variations of these
four stars 
are most probably due to starspots. 

VN01 is another W UMa binary. We cannot locate it on the CMD of the cluster
because of the lack of $B$-band data.
The star may, with equal probability, belong to NGC 362 (in which case it would lie 
on the CMD slightly above V19)  or be a field object.

If the suspected variables VN02 and VN03 are members of the cluster
then they reside on the extended 
horizontal branch (EHB). We cannot exclude such a possibility, although the physical mechanism of
the variability is unclear and the EHB of NGC 362 is rather weakly populated (Moehler et al. 2000; 
Recio-Blanco et al. 2005). 

Finally, the spotted variable VN04 may belong to NGC 362 or to the SMC. In the first case it would 
be located at the base of the red giant branch. 

\subsection{Field stars}
\label{sec:fieldvar}

In the observed field we detected 88 variables which, judging from their proper motions or CMD
location, do not belong to NGC 362. Most of these are members of the SMC whose outskirts the cluster 
is projected against. 42 objects with periods longer than 0.1 d are listed in Table \ref{tab:fieldvar}
and their light curves are shown in Figs. \ref{fig:fieldvara} -- Figs. \ref{fig:fieldvarc}. 
Several RR Lyr stars from Table \ref{tab:fieldvar} were independently discovered and cataloged by the
OGLE team (Soszy\'nski et al. 2016) and their catalog numbers are given in the last column.
All of the 46 objects with periods shorter than 0.1 d are SX Phe or $\delta$ Sct pulsators belonging 
to the SMC. They are listed in Table~\ref{tab:SXSMC} (available in its entirety from the CASE archive 
or in the electronic version of the paper). 

Stars VN05, VN06 and VN43 have ambiguous light curves, and may be classified as W UMas or ellipsoidal 
variables. 

VN07 can be phased with a period two times shorter than that given in Table~\ref{tab:SXSMC}, but such 
a period is too short for a W UMa binary (Soszy\'nski et al. 2015). The physical cause 
of the variability is most probably a reflection effect. 

VN08  is a genuine contact system (of two early B-type 
stars, judging from its color). One of the eclipses seems to be total.

The next ten entries in Table~\ref{tab:SXSMC} are RRab stars in the SMC. Three of them (VN10, VN11
and VN16) show clear Blazhko effect. In another two (VN13 and VN17) a weak Blazhko effect is possible,
but the light curves are too noisy to identify it unambiguously. 

VN19, whose proper motion indicates SMC membership, is an Algol-type system with a primary of late B-type.

The light curve of the blue star VN20 can be phased with $P=0.793313$ d given in Table~\ref{tab:SXSMC}, 
but also with $P=0.441828$ d and $P=0.306437$ d. Consequently, it is an Slowly Pulsating B-Star (SPB) pulsator. 
It is also an SMC member.

The next 13 entries are eclipsing binaries: four Algol-type systems with ellipsoidal effect (VN22, VN23, 
VN26 and VN29), and nine well detached EA systems spanning a period range of 0.80 -- 9.64 d. 

Proper motions and/or CMD locations along the SMC red giant branch suggest that stars VN34 -- VN39
are SMC members. The low amplitudes and periods from 19.5 to 29.5 d identify the first four of these 
as spotted red giants. The relatively higher amplitudes and longer periods of VN38 and VN39 suggest 
that the principal physical cause of their variability is pulsations. 

VN40 is a PM-member of the SMC. Its long period and large-amplitude variations are characteristic of 
pulsating red giants, and the very red color ($B-V=2.76$ mag) indicates it is a Mira variable. 

\begin{figure}
   \centerline{\includegraphics[width=0.95\textwidth,
               bb = 34 48 562 739, clip]{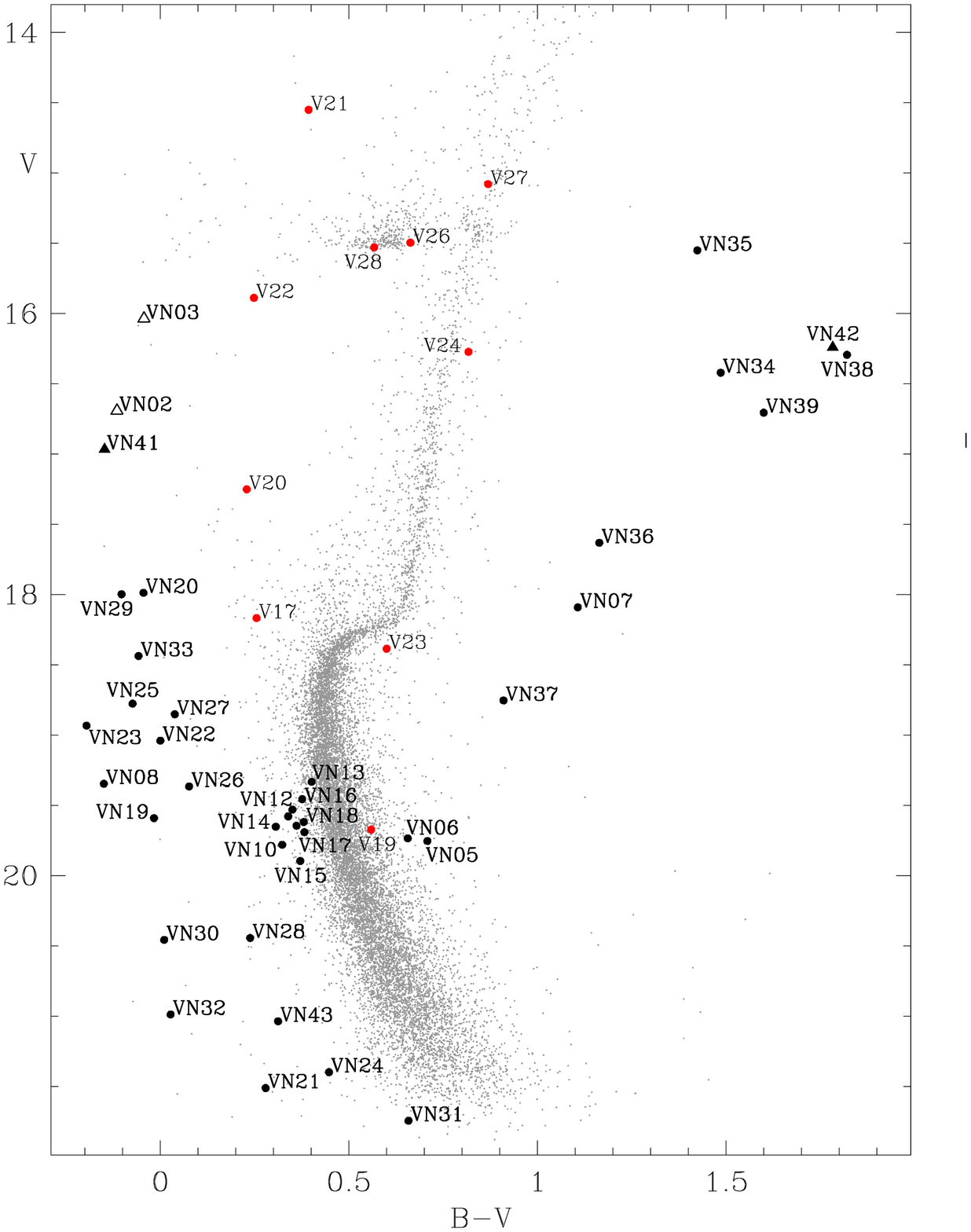}}
   \caption{CMD for the observed field with  locations of stars listed
    in Table~\ref{tab:newvar} and Table~\ref{tab:fieldvar}. Circles: variable PM-members of the cluster 
    (red) and variable field stars (black). Triangles: suspected variables (filled - field objects; open -  
    objects for which the membership data is missing or ambiguous). 
    The gray background stars are the same as in the right frame of Fig.~\ref{fig:cmds}.
    \label{fig:cmd_var}}
\end{figure}

VN41 is another PM-member of the SMC and it belongs to the young stellar population of that galaxy. Its 
color corresponds to a spectral type B3 or earlier. We did not detect any periodicity in its light
curve, which caused us to mark it as a suspected variable in Fig.~\ref{fig:cmd_var}.

VN42 is located at the tip of the red giant branch of the SMC. However, we do not have proper motion 
for this star, and we failed to detect any red-giant type variability. It is also marked as a 
suspected variable in Fig.~\ref{fig:cmd_var}.

Objects VN44 -- VN46 exhibit variations characteristic of red giants, but they are clearly too weak 
to be SMC red giants. This casts doubts on the reality of the observed variations, and 
we count them among suspected variables.

\section{New data on known variables}
\label{sec:oldvar}

We performed a cross-identification of our detected variables with the 95 stars discovered or surveyed by 
S07 and LW11, upon which the catalog of C1-12 is based. 22 of these were located beyond our FOV. 
Of the remaining stars, 26 could not be identified with any of our variables. More precisely,
we did not obtain light curves of Sz18, Sz43, LW09, LW10, and LW11, and did not detect regular
variability within $\pm$10~px from the positions of Sz10, Sz24, Sz37, Sz38, Sz42, Sz47, Sz48, 
Sz49, Sz56, Sz61, Sz62, Sz66, Sz69, Sz76, Sz77, LW04, LW05, and LW12 (all star names in this Section
and in the Appendix are taken from 
C1-12). Variability detected within $\pm$10~px from the positions of Sz41 and Sz59 turned out 
to be spurious (induced by blending with neighboring variables Sz57 in the first case, and Sz35 
in the second). Finally, the light curve of the only variable object within $\pm$10~px from 
the position of Sz25 was too poor for a detailed analysis (we are sure, however, it is not a 
Cepheid curve as suggested by S07). For the 24 variables listed in Table \ref{tab:SzLW} our 
photometry provided new information which we consider worthy of publishing. The light curves
for these objects are shown in Figs. \ref{fig:SzLWa} and \ref{fig:SzLWb}.

Sz09: The eclipses differ in depth by $\sim$0.05 mag, indicating nearly equal temperatures for 
the components. NGC~362 has $E(B-V)=0.05$ mag (H96-10). Assuming the same reddening for Sz09
(note this is a lower estimate, since the binary must be more distant from the Sun than the 
cluster), we obtain $E(B-V)_\mathrm{o}\leq-0.31$ mag, and $T_e\geq 36,000$ K characteristic of O-type stars. 
The shape of the light curve indicates that at least one component is severely tidally distorted. 
Since the light curve is smooth and stable, and no period
changes have been observed, any mass transfer between the components must be very low.
Most probably, Sz09 is a detached system whose (slightly) more massive component has just left or
is about to leave the main sequence. 

Sz12: S07 had only one night of data available  and they could not measure
a complete  light curve. We obtained a complete 
light curve of a typical RRab star, whose CMD-location clearly suggests SMC membership.

Sz17: We obtained a multiseason light curve and found that it is generally stable, with $\sim$0.05 mag 
fluctuations around the mean. The period is slightly shorter than estimated by S07 and 
stable. We confirm the star is a red giant belonging to the  cluster.

Sz19: The star is located at a projected distance $r_p\approx0.5$$r_t$ from cluster center, but both its 
CMD-location and proper motion clearly indicate that it belongs to the SMC. The multiseason light curve 
has an amplitude of nearly 1.8 mag, and a stable period of $\sim$0.5 yr. 

Sz20: This blue straggler is the only star for which we do not confirm the type of variability suggested 
by S07. According to our data it is a W UMa-type binary rather than an SX Phe pulsator. Since our coordinates 
differ by 1$''$.5 from the S07 values we cannot exclude a misidentification. However we did not find
any SX Phe-type variability within $\pm$10~px from the position of Sz20.

Sz28: For this field W UMa-type variable we provide a much better phase coverage than that of S07.

Sz31: On the archival HST frame 10615\_01\_acs\_wfc\_f435w\_drz.fits Sz31 this object splits into two 
stars of nearly equal brightness with a separation of 0$''$.7  (S07 found a separation of only 
0$''$.16, and suggested the two stars may be physically related). Upon disentangling the combined light curve 
we obtained periods of 0.53 d and 0.56 d, and derived two separate complete RRab light curves. 
The details of the analysis are given in the Appendix. Based on the CMD-location and $r_p\approx0.1$$r_t$, 
both the components of the blend are very likely members of NGC~362.

Sz32: The star, tentatively classified by S07 as EA, has in fact a nearly perfectly sinusoidal light curve
whose physical origin is difficult to establish. C1-12 counted Sz32 as a cluster member. However, 
it is too dim to be an RRc pulsator belonging to NGC~362; its location at $r_p\approx r_t$ also suggests 
that is a field star.

Sz33: S07 published an incomplete light curve. We show this cluster member to be an RRab pulsator with a 
slightly ($\sim$0.1 mag) varying amplitude, but with an otherwise stable light curve. 

Sz34: Same remarks as for Sz33, except that the amplitude of this star does not vary. The light curve 
shown in Fig. \ref{fig:SzLWa} is derived from du Pont data. Its counterpart from the Swope data has a larger 
scatter, but no amplitude modulation can be observed.

Sz36: We confirm the cluster membership of this variable red giant, as suggested by C1-12. 
The multiperiodic light curve can be 
reasonably phased with $P=238.1$ d, but also with 26.43, 39.62 and 46.50 d, the latter value being 
close to the 45 d period given by C1-12.

Sz40: For this star we observe a pronounced Blazhko effect which cannot be seen in the S07 data because of
insufficient time-coverage. Sz40 is a confirmed member of NGC~362.

Sz44: The light curve of S07 is incomplete and distorted. Our data show that Sz44 is an RRab pulsator with 
a stable light curve. With $r_p\approx0.02$$r_t$ and a CMD location among the NGC~362 RR Lyr stars it is 
a confirmed cluster member.

Sz45: Same remarks as for S44, the only difference being a weaker Blazhko effect.

Sz50: S07 published a smooth RRab-like light curve, but they remarked that it was strongly variable. 
At  first glance, our light curve in Fig. \ref{fig:SzLWb} looks somewhat chaotic. However, a careful 
analysis (detailed in the
Appendix) allowed us to conclude that Sz50 is a double-mode RRd pulsator showing the Blazhko 
effect. Like Sz44, Sz50 is a firm cluster member.

Sz51: This star, for which S07 do not provide the light curve, resides at the very tip of the red giant 
branch of the cluster. Both its proper motion and location at $r_p\approx0.03$$r_t$ strongly speak
for NGC~362 membership.

Sz55, Sz64 and Sz65: In these three cases our data show Blazhko effects of various strengths which 
were not be observed by S07 due to insufficient time coverage.

Sz68: For this star we have detected a significantly more complicated Blazhko effect than those 
reported by S07.

Sz70: S07 list this object among stars with no period and secure classification. We find it to be an 
RRab pulsator, probably with a Blazhko effect. At $V\sim$19 mag it is most likely an SMC member.

V002: Our sampling of the light curve is much denser than that of LW11; we also find the period to be 
108 d rather than 105 d. Our data indicate that V002 is a PM-member of NGC~362. 

LW03: This red giant is a cluster member. We find its period to be much longer than that 
reported by LW11 based on few measurements (174 d vs. 51 d). 

LW08: Another red giant belonging to the cluster. The dominant period seems to be 187 d rather than 63 d 
suggested by LW11. Also, we do not see their 954 d periodicity. 

\Section{Summary}
 \label{sec:sum}
We have conducted an 18-year long photometric survey of the NGC~362 field in a search 
for variable stars. A total of 100 variables plus four suspected variables were discovered, and 
multiseasonal light curves were compiled for another 46 variables that had been known before. 
Periods were obtained for all observed variables except VN41 and VN42 which appear to vary on a
timescale longer than our time base. Four new eclipsing binaries and two pulsating 
stars of SX Phe-type were found to be PM-members of NGC 362. The most interesting, but at the same 
time rather challenging objects for follow-up studies are the binary blue straggler V20 with an 
asymmetric light curve, and the 8.1 d semidetached binary V24 located at the red giant branch of 
the cluster and coinciding with a Chandra X-ray source. The latter system is a pair of $\sim$0.8 
M$_\odot$ stars: the giant primary most likely filling its Roche lobe and a slightly hotter 
subgiant secondary. The width of the secondary minimum varies by a factor of three on a timescale 
of a few years, suggesting a variable flow of matter between the components.

We also provide substantial new data on 24 of the variables cataloged by C1-12. Deserving 
further study are six RR Lyras with Blazhko effects; in particular the multimode star Sz50 whose 
lightcurve can dramatically change on a timescale of a few weeks. Finally, we collected a sample of
42 SX Phe/$\delta$ Sct pulsators belonging to the SMC.

\Acknow
{WN, WP and MR were partly supported by the grant DEC-2012/05/B/ST9/03931
from the Polish National Science Center. We thank Grzegorz Pojma\'nski for the lc code which 
vastly facilitated the work with light curves. This paper is partly based on data obtained from 
the Mikulski Archive for Space Telescopes (MAST). STScI is operated by AURA, Inc., under NASA 
contract NAS5-26555. Support for MAST for non-HST data is provided by the NASA Office of Space 
Science via grant NNX09AF08G and by other grants and contracts.
}

\section*{Appendix: The complex light curves of Sz31, Sz50 and Sz55}

\subsection{Light curve decomposition}
A first glance at light curves of some RR Lyrae stars in NGC 362 reveals evidence of multiple mode 
pulsation and/or modulation of amplitude or shape of the light curve. To investigate this behavior 
in more detail we employed a new code, Baca, for prewhitening and decomposition of a light curve 
into individual modes. In a prologue, data are normalized by subtraction from observation times 
of their median, treated as epoch of observations and by replacing observation errors with their 
inverses, treated as square root of weights. Provision exists to flag observation as invalid. Next, 
time sampling is analyzed to determine optimum frequency resolution Nyquist frequency range
from median of observation intervals. With frequency grid fixed we start core calculation by 
finding consecutive frequencies from periodogram peak  of prewhitened data. Prewhitening is 
performed at each stage by fitting the original data with the current model by Newton-Raphson 
nonlinear-least squares (NLSQ) iterations involving adjustments to amplitudes and frequencies, 
and subtracting it from the observations. Next, an AOV periodogram of prewhitened data is calculated 
by orthogonal projection of Szeg\"o trigonometric polynomials, its peak frequency 
is found by parabola fitting, and the result is appended it to the current model (Schwarzenberg-Czerny 
1996, 2012). We perform a total decomposition of the light curve by recursively repeating this 
procedure till no oscillations exceed noise. In the process, dependent frequencies are identified 
and tied to their base frequencies. In the epilogue it is possible to inspect light curves of 
individual modes by prewhitening the observations with a subset of frequencies and folding the 
result with the frequency of interest.

\subsection{Star Sz31} 
Preliminary periodograms revealed two principal frequencies. Their final values 
$f_1=1.885492451\pm0.000000054$ c/d and $f_2=1.789493550\pm0.000000055$ c/d
were obtained by prewhitening the data with a trigonometric series up to 
15 harmonics of the first frequency, and, subsequently, fitting a Szeg\"o series with 
frequency adjustment. To separate the light curves, we first converted magnitudes 
to fluxes. Next, for fixed frequencies we fitted the fluxes with the sum of two Szeg\"o 
series of frequencies $f_1$ and $f_2$ with 21 terms each. No clipping was applied. 
The final light curves plotted in the right panel of Fig. \ref{fig:Sz31} were 
obtained by adding half of the flux residuals to each model light curve, and then 
converting to magnitudes. The residuals show a mild systematic trend during the minima 
of star 2. In our opinion, this may constitute cross-talk from star 1 resulting from 
image subtraction photometry for single PSF position centered closer to star 1 than to 
star 2.

\subsection{Star Sz50} 
We analyzed separately two sets of observations containing most of the data. Each of them consisted
of two observing seasons, and spanned slightly over a year: Set 1 with 1028 points from HJD 3917 till 
HJD 4362, and Set 2 with 422 points from 
HJD 4698 till HJD 4699. The two strongest oscillations in Set 1 have amplitudes of 0.117 and 0.083 mag, 
and frequencies $f_0=2.0483\pm 0.0010$ and $f_1=2.8048\pm 0.0025$ c/d, respectively. In Set 2 the 
strongest peak corresponds to $2f_0$ harmonic of amplitude 0.083 mag, while the second and third ones 
with amplitudes of 0.066 and 0.051 mag - to $f_0+1$ c/d alias, and to $f_1$, respectively.  
The ratio $f_0/f_1=0.73$ is consistent with Petersen diagrams of fundamental and first overtone 
bimodal RR Lyr (see e.g. Moskalik 2013). 
The frequencies and their errors quoted above were derived from the analysis of harmonic and 
combination peaks at higher frequencies corresponding to signatures [1,0], [0,1], [2,0], [1,0], [1,1],
[3,0], [2,1], [3,1], [4,0] in Set 1 and [2,0], [1,0], [0,1], [1,1], [4,0] in Set 2, where
the first and second digit is harmonic and base frequency number, respectively. Solving by 
least squares for Set 1 and Set 2 independently, we obtained two values of $f_0$ and two of $f_1$  
(in each case consistent within less than $1.5\sigma$) which were subsequently averaged. 
A cautionary remark is due here: given that no season spans 
more than 70 days, we cannot completely exclude year aliasses $\pm$0.0027 c/d of base frequencies. 
However, base frequencies derived from harmonics are less affected. 

\begin{figure}
   \centerline{\includegraphics[width=0.95\textwidth,
       bb = 70 537 510 765, clip]{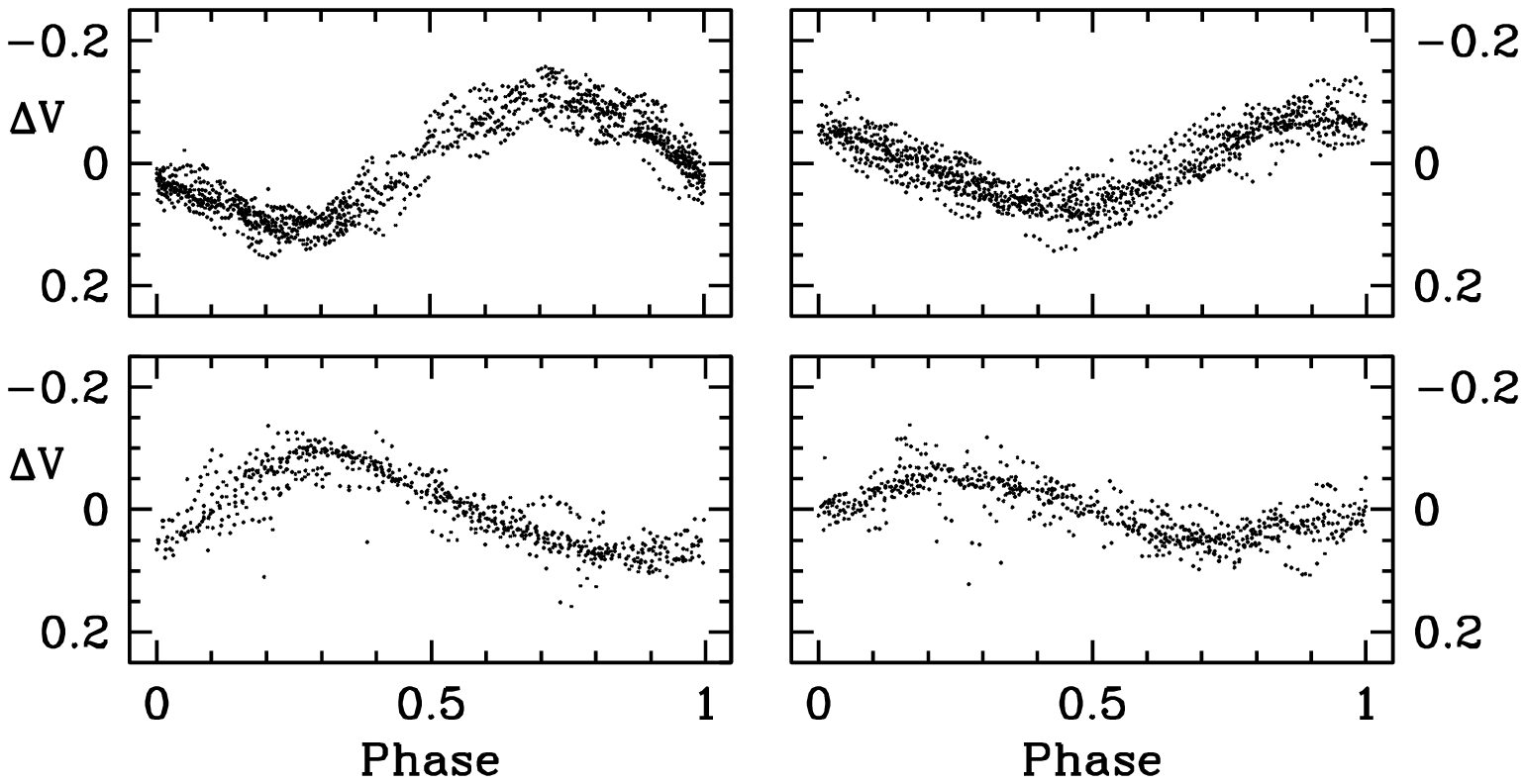}}
   \caption{The light curve of Sz50 phased with frequencies $f_0$ and $f_1$ for Set~1 (upper
    left and right, respectively), and $2f_0$ and $f_1$ for Set 2 (lower left and right, respectively).
    In each case all other modes have been prewhitened.
    \label{fig:Sz50}}
\end{figure}

Two more facts are noteworthy. First, at amplitudes of 0.075 and 0.083 mag the $2f_0$ harmonic is strong 
in Set 1 and dominant in Set 2, making an impression of period halving. Second, in Set 1 there is evidence 
of a weak oscillation with amplitude 0.019 mag and frequency 1.0120 c/d - a possible subharmonic 
$f_0/2$. 
The amplitude of this sub-harmonic in the whole data is even stronger, at 0.030 mag.
Additionally, our data yield strong evidence of amplitude modulation of principal modes on time scales
over 100 days. In four panels of Fig. \ref{fig:Sz50} we plot light curves phased with 
frequencies $f_0$ and $f_1$ for Set 1, and $2f_0$ and $f_1$ for Set 2 after prewhitening  
of all other modes. Appreciable changes of their amplitude and shape may indicate a Blazhko modulation. 
A further hint towards the Blazhko modulation is our inability to remove power near $f_0$ by 
fitting just one sinusoid to Set 1 data: two components spaced by 0.010 c/d were needed, whereas in case 
of window-function ghosts removing the stronger component usually removes the fainter ones, too. 
Similarly, three components were needed to remove $f_0$ and two components to remove $f_1$ from 
the whole dataset. Thus, there is some evidence of a long-term frequency modulation.

Apart from that Set 1 yields no unexplained oscillations with amplitudes exceeding 0.01 mag.
In Set 2, amplitudes of $\approx 0.02$ mag appear at frequencies 1.77
and 0.69 c/d, close to $f_0/3$. However this data set is poorer, consisting of runs not exceeding 
40 d, hence we refrain from further discussion. 

\subsection{Star Sz55} 
Analysis of this star reveals presence of one pulsation mode at a frequency $f_0=1.96185\pm 
41$ c/d together with its multiple harmonics and their day aliasses. Apart from that, up to frequency 
of 35 c/d no unrelated oscillations with amplitudes over 0.02 mag can be detected.
Similarly as in the case of Sz50, we derived the final value of the period by averaging results from 
three data sets from HJD ranges given in Table \ref{tab:ampl_changes}.
Actually, for this purpose we used $5f_0$ harmonics present in all data sets, as their relative error 
was a factor of $\sim$5 less than that for $f_0$ itself. 
Assuming a 
coherence throughout the whole data set would be risky, given pronounced changes of the shape of 
the light curve. However, the consistency of results 
speaks for period stability and correct cycle count, while the scatter of individual 
values yields a robust estimate of the error. 

\begin{figure}
   \centerline{\includegraphics[width=0.95\textwidth,
       bb = 44 545 561 740]{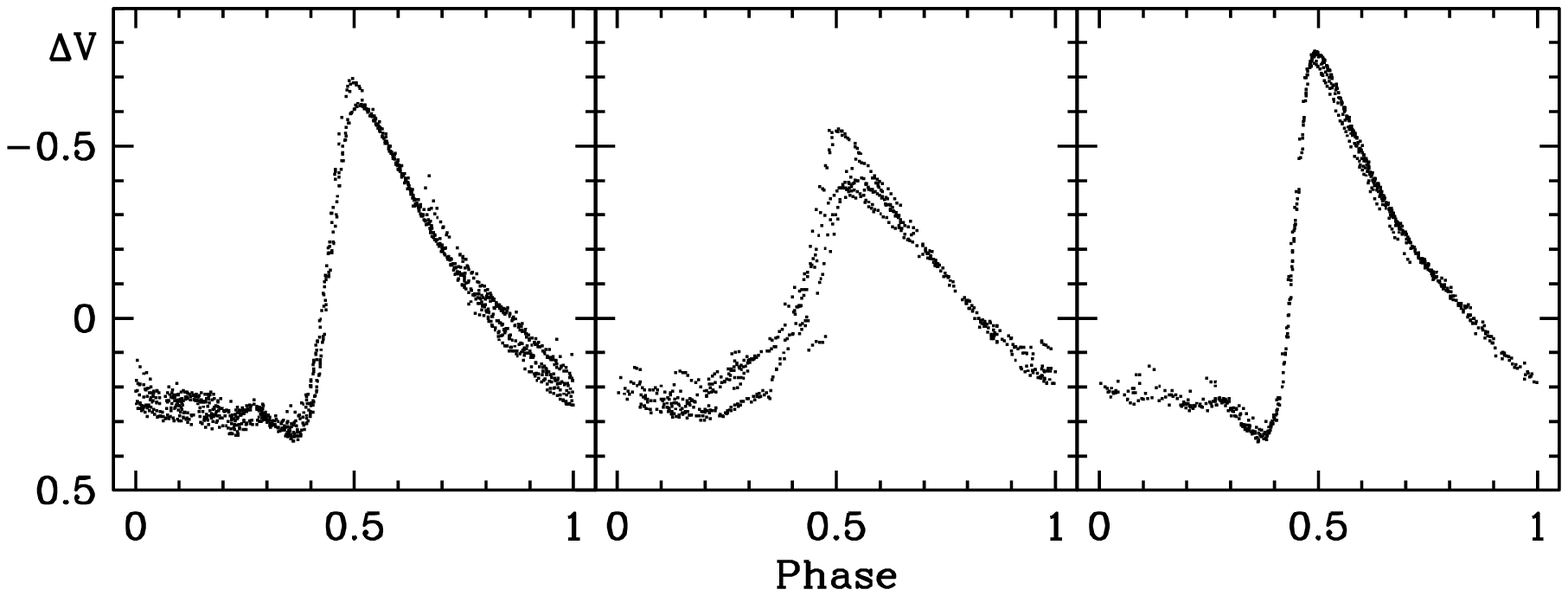}}
   \caption{Seasonal changes of shape and amplitude of the Sz55 light curve. From left to right, 
    the panels correspond to HJD ranges defined in Table \ref{tab:ampl_changes}. All curves are 
    phased with the frequency $f_0$.
    \label{fig:Sz55}}
\end{figure}

\noindent Up to 10 harmonics reveal amplitudes exceeding 0.01 mag, though on occasions the strongest peaks 
correspond to their $\pm 1$ and $\pm 2$ c/d aliasses. This said, it must be stressed that their 
amplitudes are anything but stable, revealing between seasons changes of up to factor 2 both in 
absolute and relative sense. The corresponding changes of light curve shape are illustrated in three 
panels of Fig. \ref{fig:Sz55}. However, the time scale of this modulation likely is of a large 
fraction of a year (otherwise the peaks in our periodogram would reveal broadening or subcomponents). 
Due to large seasonal gaps we are unable to check this long-term modulation
for periodicity, but a first guess would be we are observing Blazhko-type modulation on a time 
scale of a few hundred days. 

\clearpage

\begin{table}[H]
\footnotesize
 \begin{center}
 \caption{\footnotesize Basic data of newly discovered variables belonging or likely 
          belonging to NGC 362}
          \label{tab:newvar}
 \begin{tabular}{|l|c|c|c|c|c|c|c|c|}
  \hline
 ID & RA & DEC & $V$ & $B-V$ & $\Delta_V$ &Period & Type$^a$ & Mem$^c$\\
    & [deg] & [deg]  &[mag]& [mag] & [mag] & [d] & Remarks$^b$ & \\
  \hline
V17 & 15.695623 & -70.762892 & 18.168 & 0.255 & 0.015 & 0.034562 & SX; BS & Y\\
V18 & 15.825319 & -70.842839 & 17.671 &  & 0.451 & 0.067906 & SX & Y\\
V19 & 15.639185 & -70.784220 & 19.673 & 0.559 & 0.449 & 0.231436 & EW & Y\\     
V20 & 15.831187 & -70.844957 & 17.251 & 0.229 & 0.737 & 0.402175 & EB/EW; BS & Y\\  
V21 & 15.819295 & -70.847854 & 14.550 & 0.393 & 0.460 & 0.520397 & RRab & Y\\   
V22 & 15.801621 & -70.846397 & 15.888 & 0.248 & 0.236 & 0.566332 & RRab & Y\\   
V23 & 15.732048 & -70.852705 & 18.385 & 0.600 & 0.210 & 0.920841 & EA & Y\\     
V24 & 15.842032 & -70.845761 & 16.272 & 0.817 & 0.266 & 8.140462 & EA; RG; X & Y\\  
V25 & 15.893542 & -70.899521 & 15.727 &	      & 0.681 & 12.51458 & Sp & Y\\
V26 & 15.659105 & -70.914325 & 15.495 & 0.663 & 0.020 & 14.72401 & Sp & Y\\     
V27 & 15.875219 & -70.922811 & 15.079 & 0.869 & 0.091 & 15.05222 & Sp & Y\\     
V28 & 15.669002 & -70.816794 & 15.529 & 0.567 & 0.028 & 24.30491 & Sp & Y\\     
VN01 & 15.767435 & -70.841165 & 18.811 &  & 0.840 & 0.261285 & EW & U\\
VN02 & 15.567987 & -70.861797 & 16.692 & -0.115 & 0.111 & 0.764585 & $s$ & U\\    
VN03 & 15.698946 & -70.668651 & 16.037 & -0.043 & 0.131 & 1.081190 & $s$ & U\\    
VN04 & 15.485683 & -70.963615 & 18.255 &  & 0.247 & 13.45729 & Sp & U\\
  \hline
 \end{tabular}
\end{center}
{\footnotesize
$^a$ EA - detached eclipsing binary, EB - close eclipsing binary, EW - contact binary, 
Ell - Ellipsoidal variable, Sp - spotted variable, RRab - RRab pulsator, SX - SX Phe pulsator,
$s$ - suspected variable.\\
$^b$ BS- blue straggler, RG - red giant, X - X-ray source. \\
$^c$ Membership status: Y - member, U - no data or data ambiguous.}
\end{table}

\clearpage

\begin{table}[H]
\footnotesize
 \begin{center}
 \caption{\footnotesize Basic data of field variables identified within the present survey}
          \label{tab:fieldvar}
 \begin{tabular}{|l|c|c|c|c|c|c|c|c|}
  \hline
 ID & RA & DEC & $V$ & $B-V$ & $\Delta_V$ &Period & Type$^a$ & OGLE\\
    & [deg] & [deg]  &[mag]& [mag] & [mag] & [d] &  & ID$^b$ \\
  \hline
VN05 & 15.835164 & -70.821843 & 19.754 & 0.708 & 0.268 & 0.229394 & EW & \\
VN06 & 15.701709 & -70.840557 & 19.735 & 0.656 & 0.362 & 0.261982 & EW & \\
VN07 & 16.101429 & -70.757218 & 18.091 & 1.107 & 0.055 & 0.337148 & Ell? & \\
VN08 & 15.905190 & -70.672751 & 19.346 & -0.150 & 0.632 & 0.472937 & EW & \\
VN09 & 15.926041 & -70.758607 & 19.579 & 0.339 & 0.937 & 0.547801 & RRab & 4642\\
VN10 & 15.508883 & -70.840514 & 19.781 & 0.323 & 0.845 & 0.595133 & RRab & 4560\\
VN11 & 15.835365 & -70.804176 & 19.645 & 0.361 & 0.779 & 0.617199 & RRab & 4625\\
VN12 & 16.146925 & -70.924962 & 19.531 & 0.350 & 0.644 & 0.620290 & RRab & 4669\\
VN13 & 15.752016 & -70.808638 & 19.333 & 0.401 & 0.485 & 0.624751 & RRab & 4601\\
VN14 & 15.789838 & -70.755247 & 19.652 & 0.306 & 0.584 & 0.631775 & RRab & 4614\\
VN15 & 15.745504 & -70.661926 & 19.896 & 0.371 & 0.531 & 0.634066 & RRab & 4599\\
VN16 & 15.541113 & -70.976611 & 19.456 & 0.376 & 0.429 & 0.648357 & RRab & \\
VN17 & 16.169725 & -70.819698 & 19.691 & 0.382 & 0.505 & 0.650213 & RRab & 4672\\
VN18 & 16.189086 & -70.870573 & 19.619 & 0.380 & 0.671 & 0.689559 & RRab & 4673\\
VN19 & 15.472769 & -70.890452 & 19.591 & -0.017 & 0.658 & 0.707474 & EA & \\
VN20 & 15.645727 & -70.920385 & 17.988 & -0.045 & 0.103 & 0.793313 & SPB & \\
VN21 & 15.552856 & -70.966519 & 21.512 & 0.279 & 0.960 & 0.795833 & EA & \\
VN22 & 15.805797 & -70.871584 & 19.040 & 0.000 & 0.247 & 0.874733 & EA & \\
VN23 & 15.660924 & -70.836421 & 18.933 & -0.196 & 0.440 & 1.039676 & EA & \\
VN24 & 15.749289 & -70.776986 & 21.399 & 0.447 & 0.833 & 1.083923 & EA & \\
VN25 & 15.784002 & -70.680881 & 18.776 & -0.074 & 0.321 & 1.213690 & EA & \\
VN26 & 16.008099 & -71.015307 & 19.366 & 0.076 & 0.235 & 1.358603 & EA & \\
VN27 & 15.698910 & -70.823079 & 18.852 & 0.038 & 0.161 & 3.099988 & EA & \\
VN28 & 15.463015 & -70.889363 & 20.444 & 0.238 & 0.616 & 3.576705 & EA & \\
VN29 & 15.849144 & -70.959040 & 17.998 & -0.103 & 0.100 & 4.110868 & EA & \\
VN30 & 15.876971 & -70.675557 & 20.458 & 0.010 & 0.488 & 4.400843 & EA & \\
VN31 & 16.047540 & -70.756549 & 21.745 & 0.658 & 1.033 & 4.415142 & EA & \\
VN32 & 15.746235 & -70.900027 & 20.989 & 0.027 & 0.788 & 5.782922 & EA & \\
VN33 & 15.729061 & -71.027554 & 18.437 & -0.058 & 0.198 & 9.640647 & EA & \\
VN34 & 15.545848 & -70.944889 & 16.421 & 1.486 & 0.054 & 15.05222 & Sp; RG & \\
VN35 & 15.737894 & -70.984428 & 15.550 & 1.424 & 0.029 & 19.46125 & Sp; RG & \\
VN36 & 15.819203 & -70.922276 & 17.632 & 1.164 & 0.058 & 24.30491 & Sp; RG & \\
VN37 & 16.041693 & -70.753661 & 18.754 & 0.910 & 0.090 & 29.54246 & Sp; RG & \\
VN38 & 15.692691 & -70.677675 & 16.294 & 1.821 & 0.249 & 55.94843 & Sp?; RG & \\
VN39 & 16.019315 & -70.945663 & 16.705 & 1.600 & 0.184 & 62.95116 & Sp?; RG & \\
VN40 & 15.956149 & -70.730158 & 16.950 & 2.764 & 1.060 & 262.2845 & LPV; RG & \\
VN41 & 15.912747 & -70.819997 & 16.966 & -0.148 & 0.167 & 4500.000$^c$ & $s$ & \\
VN42 & 16.035620 & -70.976718 & 16.240 & 1.783 & 0.285 & 4600.000$^c$ & RG; $s$ & \\
VN43 & 16.170235 & -70.884000 & 21.037 & 0.312 & 0.434 & 0.172275 & EW & \\
VN44 & 16.151624 & -70.732550 & 22.301 &       & 0.553 & 18.71347 & $s$ & \\
VN45 & 15.577582 & -70.876860 & 22.829 &       & 1.073 & 27.33038 & $s$ & \\
VN46 & 15.833286 & -70.905513 & 21.149 &       & 1.110 & 107.8565 & $s$ & \\
  \hline
 \end{tabular}
\end{center}
{\footnotesize
$^a$ EA - detached eclipsing binary, EB - close eclipsing binary, EW - contact binary,
Ell - Ellipsoidal variable, Sp - spotted variable, RRab - RRab pulsator, SPB - SPB pulsator,
LPV - long period variable, RG - red giant, $s$ - suspected variable.\\
$^b$ From Soszy\'nski et al. (2016).\\
$^c$ Time-span between the first and the last observation.}
\end{table}

\begin{table}[H]
\footnotesize
 \begin{center}
 \caption{\footnotesize Basic data of 46 SX Phe/$\delta$ Sct pulsators from the SMC
          identified within the present survey$^*$}
          \label{tab:SXSMC}
 \begin{tabular}{|l|c|c|c|c|c|}
  \hline
    ID & RA & DEC & $V$ & $B-V$ &Period \\
       & [deg] & [deg]  &[mag]& [mag]  & [d] \\
  \hline
VN47 & 16.179903 & -71.005473 & 22.967 &       & 0.031072\\
VN48 & 16.075408 & -70.925832 & 22.309 & 0.122 & 0.039332\\
... & ... & ... & ... & ... & ...\\
VN91 & 15.638799 & -70.962703 & 21.615 & 0.239 & 0.074687\\
VN92 & 16.124508 & -70.804743 & 21.767 & 0.118 & 0.077352\\
  \hline
 \end{tabular}
\end{center}
{\footnotesize
$^*$ Available in its entirety from the CASE archive or in the electronic version of the paper. }
\end{table}

\begin{table}[H]
\footnotesize
 \begin{center}
 \caption{\footnotesize Basic data of S07 and LW11 variables for which important new 
          information is provided}
          \label{tab:SzLW}
 \begin{tabular}{|l|c|c|c|c|c|c|c|c|}
  \hline
 ID$^a$ & RA & DEC & $V$ & $B-V$ & $\Delta_V$ &Period & Type$^b$ & Mem$^d$\\ 
    & [deg] & [deg]  &[mag]& [mag] & [mag]    & [d]     & Remarks$^c$ & \\
  \hline
Sz09 & 15.491038 & -70.991094 & 15.837 & -0.262 & 0.270 & 1.252792 & EA & N\\
Sz12 & 15.500430 & -70.983710 & 19.686 & 0.351 & 0.589 & 0.697098 & RRab & N\\  
Sz17 & 15.658535 & -70.856023 & 13.391 & 1.232 & 0.156 & 67.73800 & LPV; RG & Y\\
Sz19 & 15.666372 & -70.774557 & 17.343 & 3.437 & 1.796 & 187.3436 & LPV & N\\
Sz20 & 15.675416 & -70.783423 & 17.567 & 0.147 & 0.124 & 0.356646 & EW; BS & Y\\    
Sz28 & 15.747879 & -70.987791 & 18.969 & 0.825 & 0.535 & 0.238010 & EW & N\\    
Sz31 & 15.768227 & -70.855610 & 14.751 & 0.310 & 0.707 & 0.530364 & RRab & Y\\  
Sz32 & 15.771615 & -71.001746 & 17.045 & 0.263 & 0.148 & 0.585562 & Ell? & U\\  
Sz33 & 15.772487 & -70.850987 & 15.540 & 0.268 & 0.613 & 0.644707 & RRab; Bl & Y\\  
Sz34 & 15.780658 & -70.848124 & 15.414 & 0.178 & 0.549 & 0.643308 & RRab & Y\\  
Sz36 & 15.782375 & -70.829589 & 12.766 & 1.415 & 0.164 & 237.7636 & LPV; RG & Y\\
Sz40 & 15.788274 & -70.866926 & 15.535 & 0.322 & 0.983 & 0.517682 & RRab; Bl & Y\\  
Sz44 & 15.798804 & -70.846290 & 15.206 & 0.266 & 0.751 & 0.549218 & RRab & Y\\  
Sz45 & 15.799041 & -70.847472 & 15.005 & 0.316 & 0.785 & 0.514701 & RRab; Bl & U\\  
Sz50 & 15.810568 & -70.864042 & 15.245 & 0.410 & 0.373 & 0.489414 & RRd; Bl & Y\\  
Sz51 & 15.812784 & -70.842350 & 11.695 & 2.187 & 0.758 & 138.0398 & LPV; RG & Y\\
Sz55 & 15.823751 & -70.838109 & 15.348 & 0.391 & 0.921 & 0.509843 & RRab; Bl & Y\\  
Sz64 & 15.845882 & -70.843214 & 15.327 & 0.298 & 0.634 & 0.607315 & RRab; Bl & Y\\  
Sz65 & 15.851014 & -70.845423 & 15.417 & 0.414 & 0.268 & 0.684739 & RRab; Bl & Y\\  
Sz68 & 15.886111 & -70.888826 & 15.492 & 0.331 & 0.917 & 0.474424 & RRab; Bl & Y\\  
Sz70 & 15.897164 & -70.908059 & 19.025 & 0.456 & 0.633 & 0.583591 & RRab & N\\  
V002 & 15.840947 & -70.905584 & 12.835 & 1.522 & 1.220 & 107.5510 & LPV; RG & Y\\
LW03 & 15.806756 & -70.843614 & 12.897 & 1.395 & 0.749 & 173.5635 & LPV; RG & Y\\
LW08 & 15.887487 & -70.827051 & 12.645 & 1.561 & 0.242 & 186.8250 & LPV; RG & Y\\
  \hline
 \end{tabular}
\end{center}
{\footnotesize 
$^a$ After C1-12.\\
$^b$ EA - detached eclipsing binary, EW - contact binary, Ell - Ellipsoidal variable, 
RRab - RRab pulsator, RRd - RRd pulsator.\\
$^c$ BS- blue straggler, RG - red giant, Bl - Blazhko effect. \\
$^d$ Membership status: Y - member, N - nonmember, U - no data or data ambiguous.}
\end{table}

\begin{table}[H]
\footnotesize
 \begin{center}
 \caption{Seasonal change of amplitudes in Sz55}
        \label{tab:ampl_changes}
 \begin{tabular}[]{clll}
  \hline
%~~\verb|\|~ Set: & 0 & 1 & 2 \\
& \multicolumn{3}{c}{Ranges of (HJD-2450000)}\\
Freq.&{1755--2135} &{3918--4362} & {4699--5128} \\
  \hline
$f_0$ & 0.520 & 0.378 & 0.319 \\
2$f_0$ & 0.187 & 0.163 & 0.093 \\
3$f_0$ & 0.226 & 0.112 & 0.054 \\
4$f_0$ & 0.087 & 0.064 & 0.038$^a$ \\
5$f_0$ & 0.116 & 0.050 & 0.026 \\ 
\hline
\multicolumn{4}{p{7.0 cm}}
{\footnotesize $^a$ $4f_0-1$ alias}
\end{tabular}
\end{center}
\end{table}
 
\begin{subfigures}
\begin{figure}
   \centerline{\includegraphics[width=0.95\textwidth,
               bb = 36 22 528 767, clip]{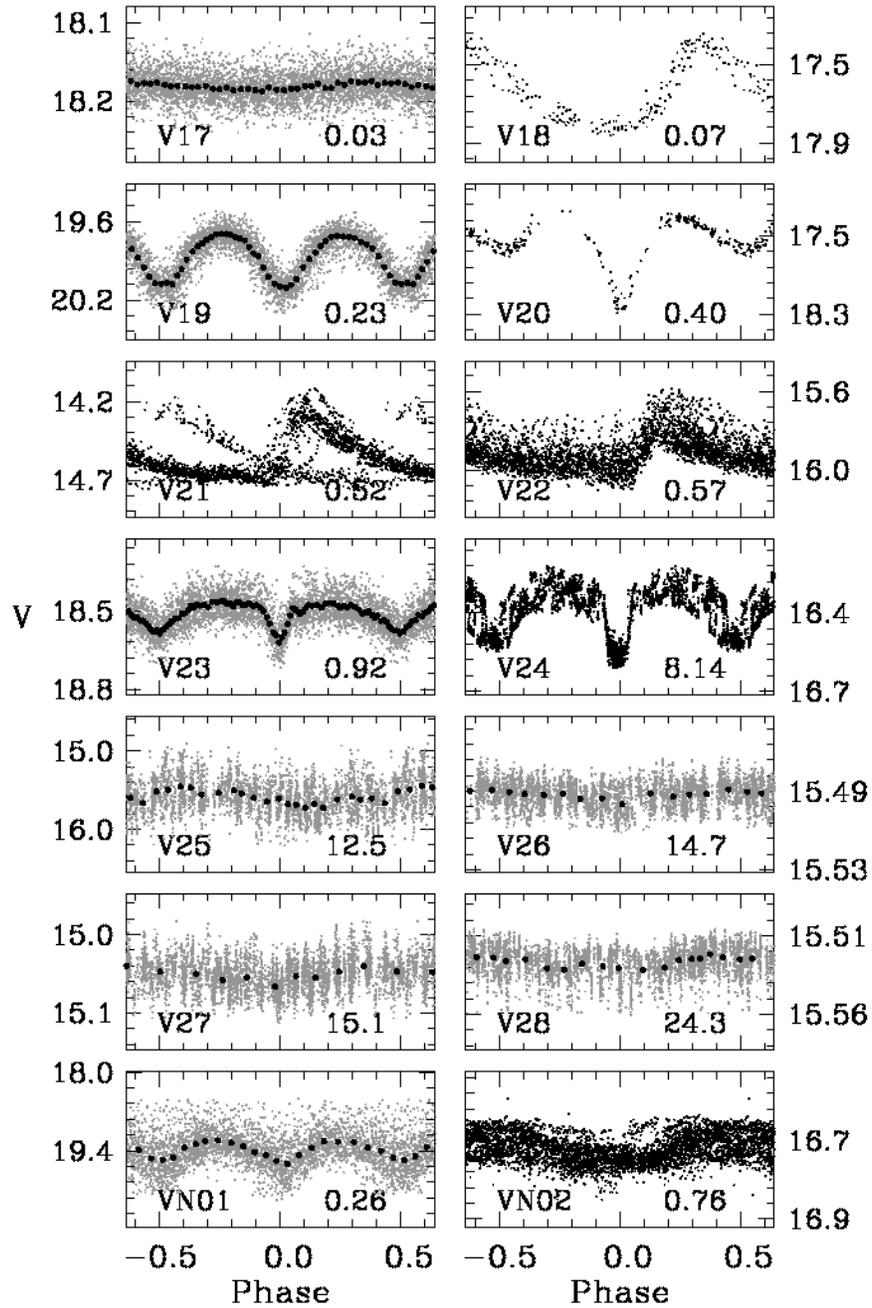}}
   \caption{Phased $V$-band light curves of the newly detected variables which are members 
    or likely members of NGC~362. Phase-binned data are shown for selected stars with 
    heavy black points. Individual panel labels give star ID and period in days.
    \label{fig:newvara}}
\end{figure}

\begin{figure}
   \centerline{\includegraphics[width=0.95\textwidth,
       bb = 42 629 528 765, clip]{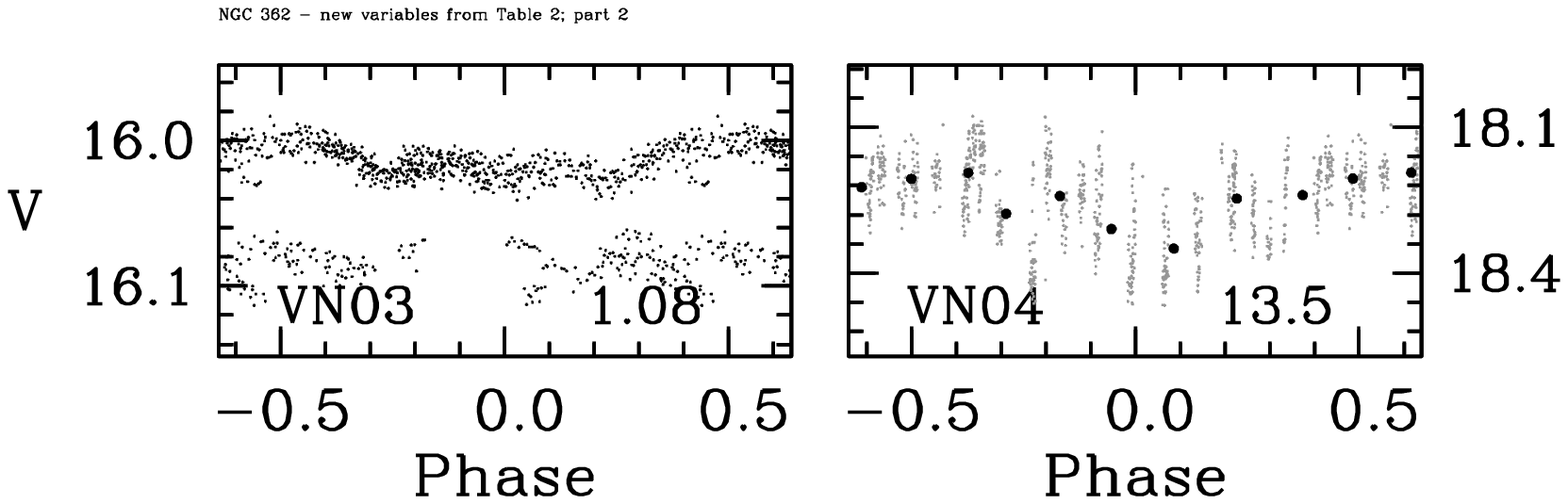}}
   \caption{Continuation of Fig. \ref{fig:newvara}.  
    \label{fig:newvarb}}
\end{figure}
\end{subfigures}

\begin{subfigures}
\begin{figure}
   \centerline{\includegraphics[width=0.95\textwidth,
               bb = 36 22 528 767, clip]{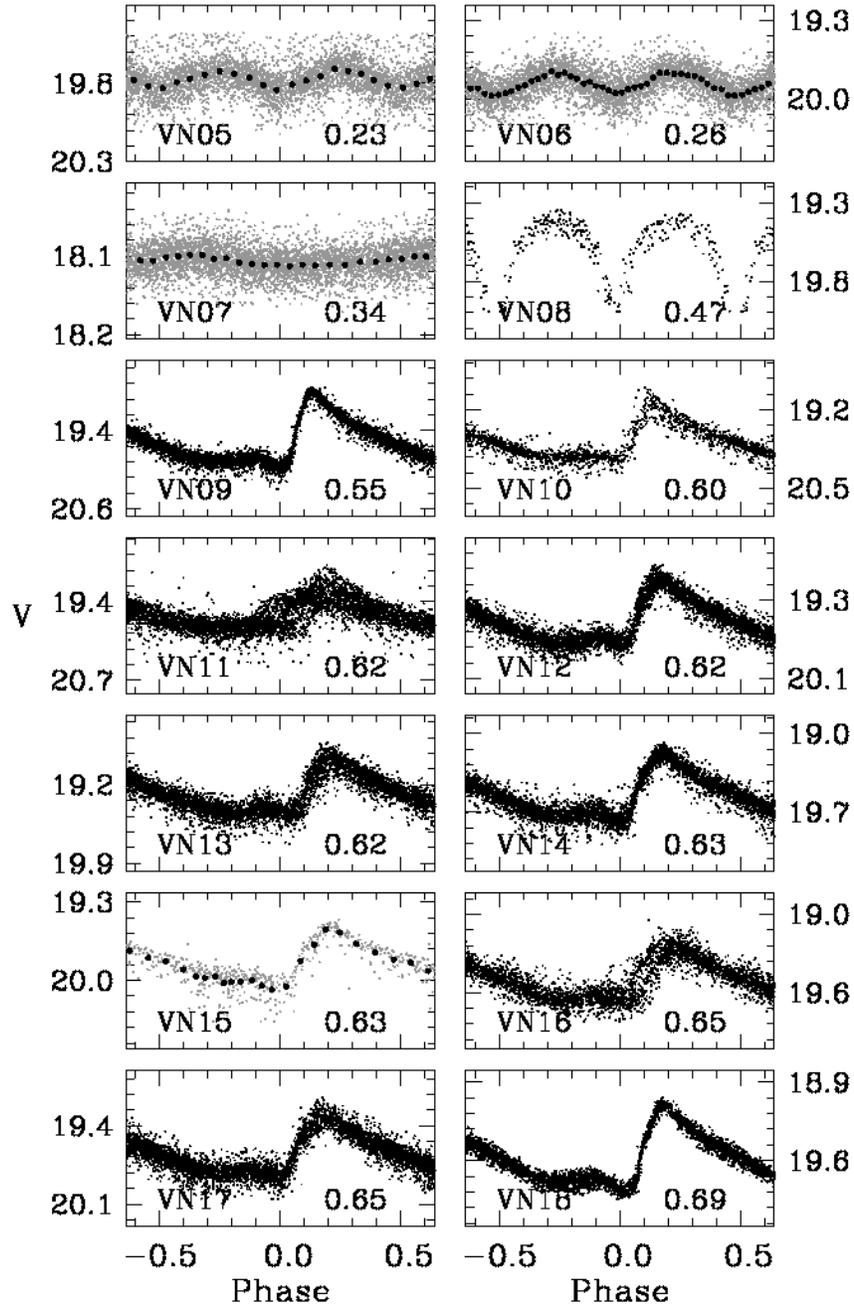}}
   \caption{Phased $V$-band light curves of stars listed in Table~\ref{tab:fieldvar}.
    Phase-binned data are shown for selected stars with heavy 
    black points. Individual panel labels give star ID and period in days.
    \label{fig:fieldvara}}
\end{figure}

\begin{figure}
   \centerline{\includegraphics[width=0.95\textwidth,
       bb = 36 22 528 765, clip]{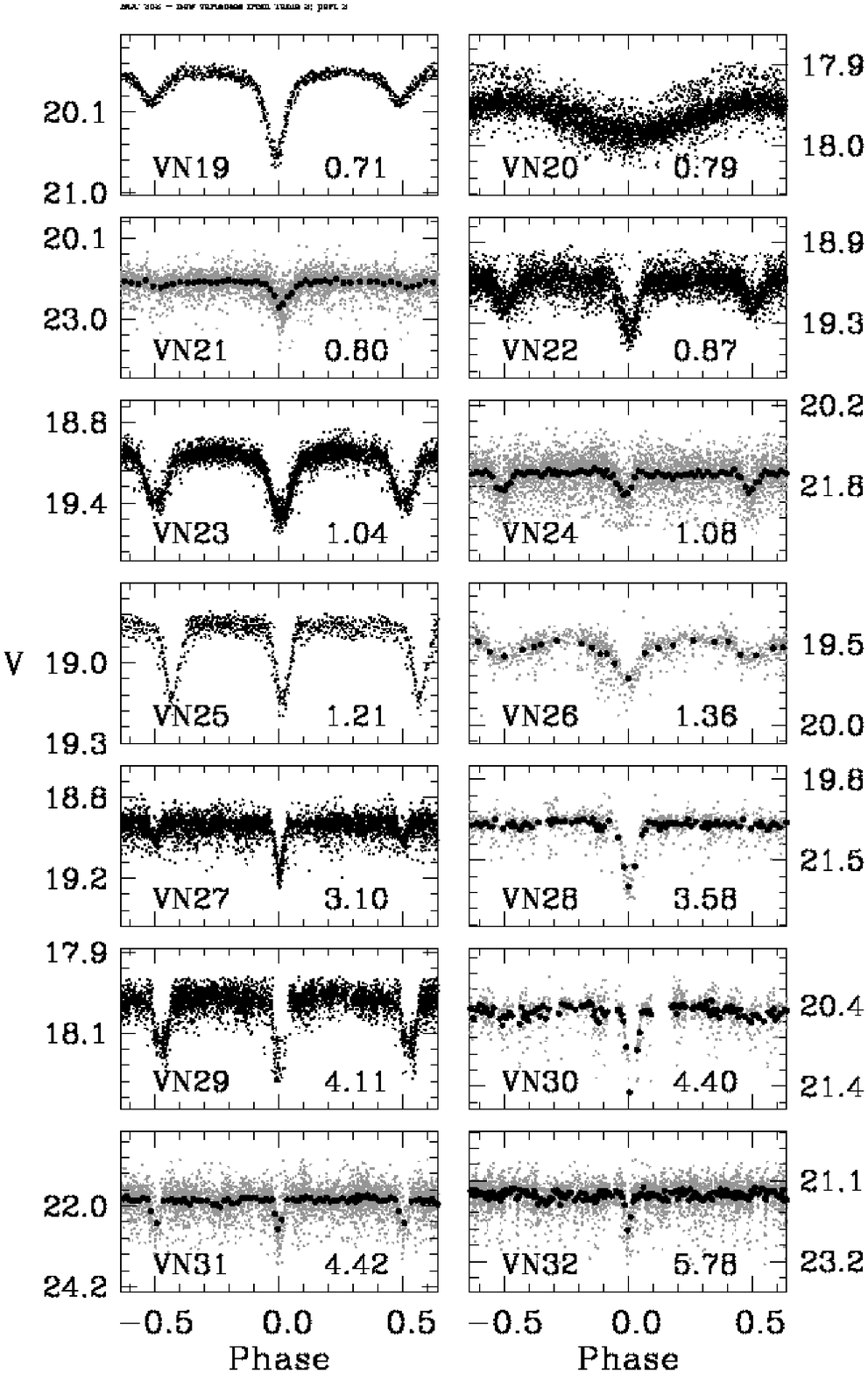}}
   \caption{Continuation of Fig. \ref{fig:fieldvara}.  
    \label{fig:fieldvarb}}
\end{figure}

\begin{figure}
   \centerline{\includegraphics[width=0.95\textwidth,
       bb = 36 22 528 765, clip]{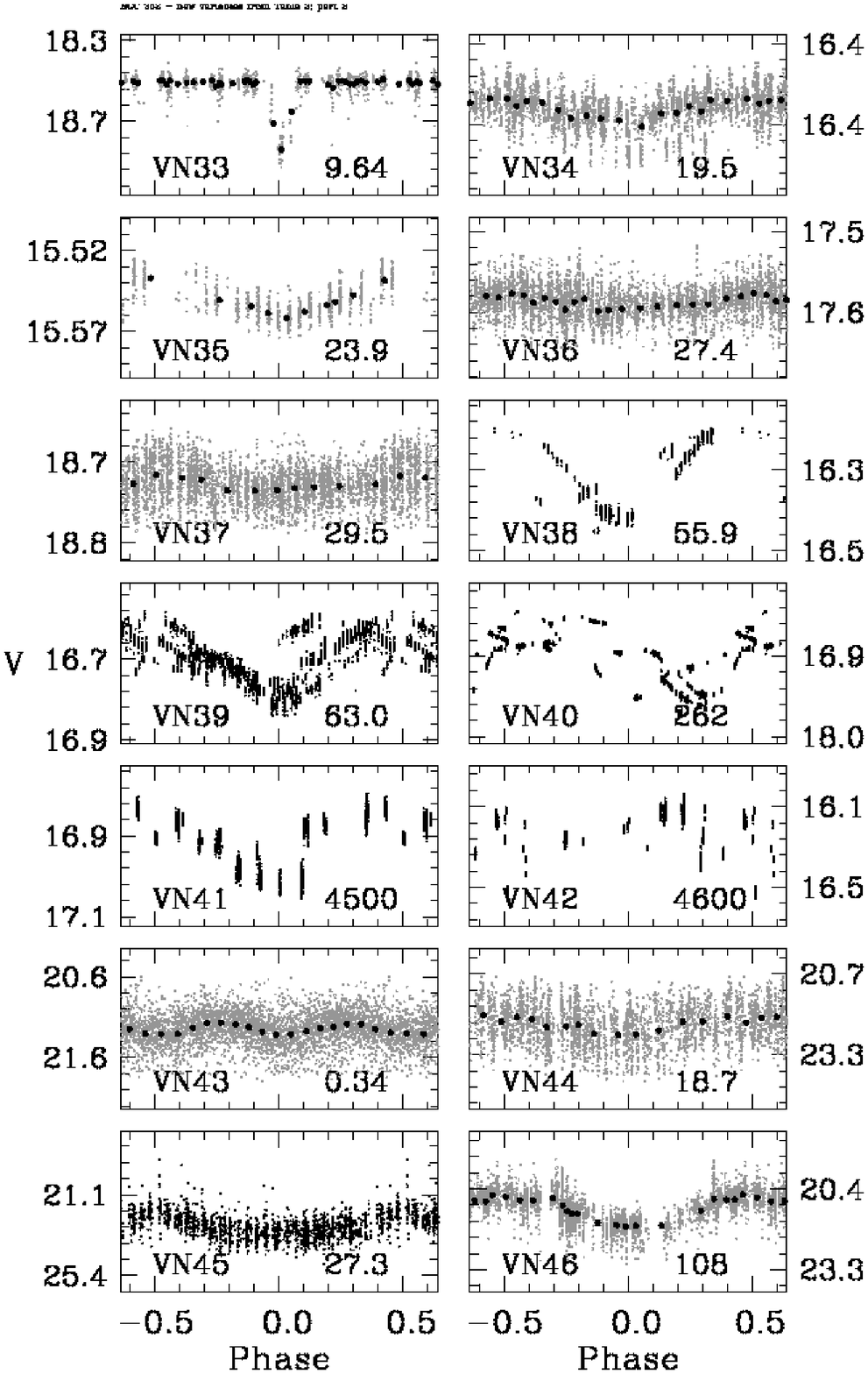}}
   \caption{Continuation of Fig. \ref{fig:fieldvarb}.  
    \label{fig:fieldvarc}}
\end{figure}
\end{subfigures}

\begin{subfigures}
\begin{figure}
   \centerline{\includegraphics[width=0.95\textwidth,
               bb = 42 26 528 767, clip]{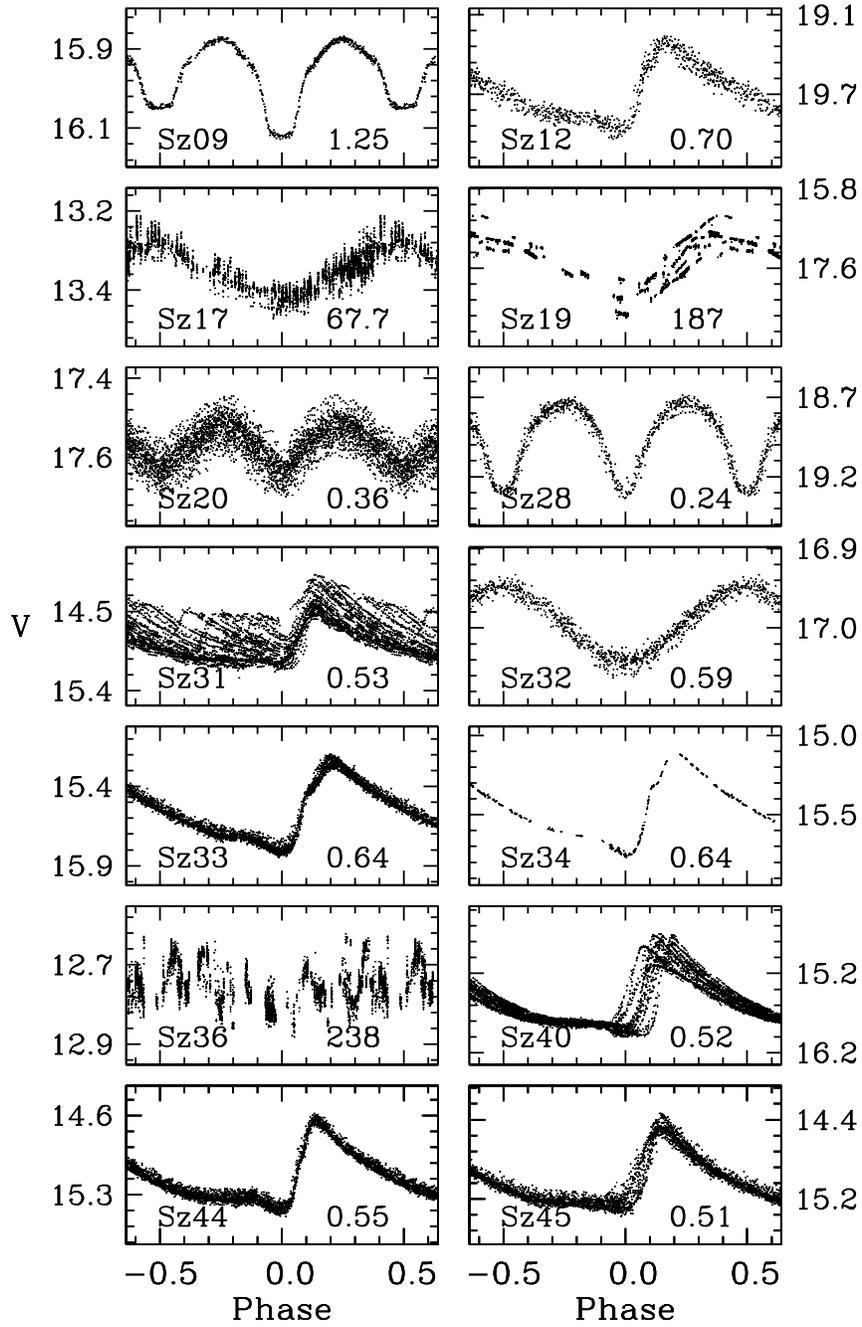}}
   \caption{Phased $V$-ligt curves of the S07 and LW11 variables listed in 
    Table~\ref{tab:SzLW}. 
    Inserted labels give star ID and period in days.
    \label{fig:SzLWa}}
\end{figure}

\begin{figure}
   \centerline{\includegraphics[width=0.95\textwidth,
       bb = 42 226 528 765, clip]{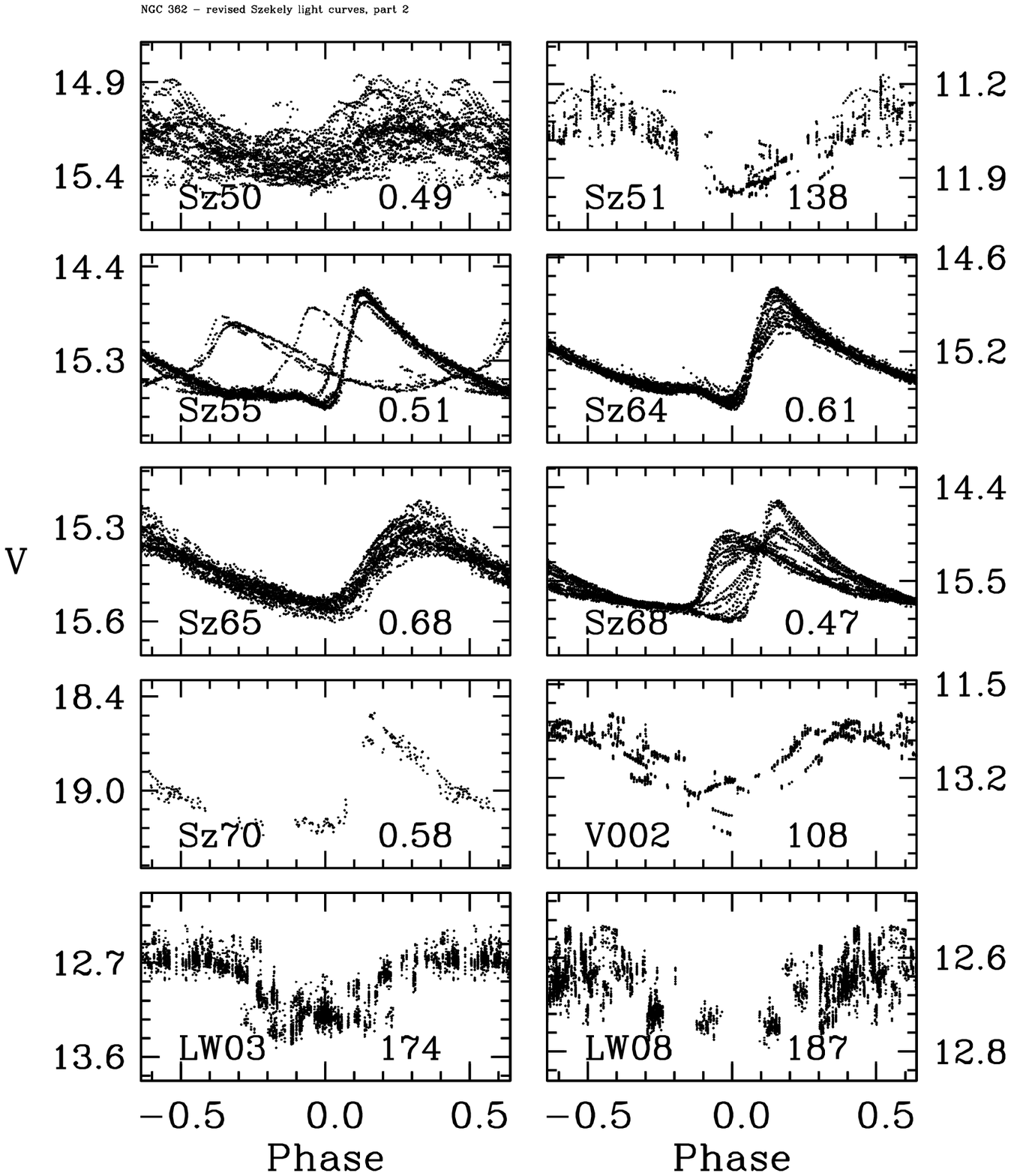}}
   \caption{Continuation of Fig. \ref{fig:SzLWa}.  
    \label{fig:SzLWb}}
\end{figure}
\end{subfigures}

\begin{figure}
   \centerline{\includegraphics[width=0.95\textwidth,
       bb = 70 640 510 765, clip]{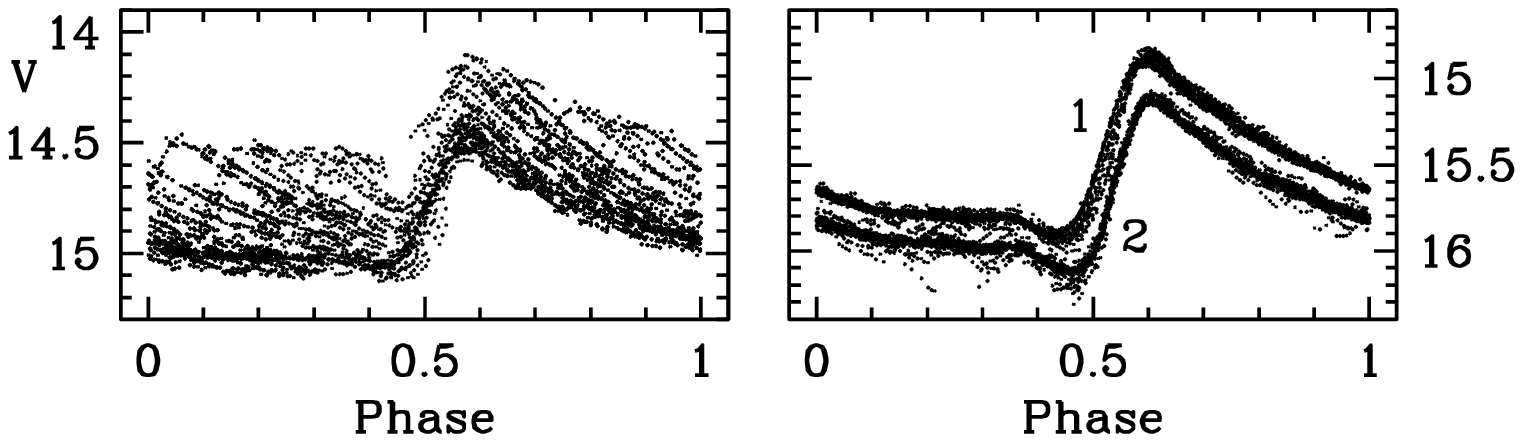}}
   \caption{Left: the observed light curve of the Sz31 blend phased with $P=0.530364$ d 
    (same as in Fig. \ref{fig:SzLWa}). Right: disentangled light curves of the components of the blend 
     phased with $P_1=0.530366$ d and $P_2=0.558817$ d. The curve \#2 is shifted down by 
     0.2 mag for clarity.  
    \label{fig:Sz31}}
\end{figure}

\end{document}